\documentclass{ieeeaccess}
\usepackage{cite}
\usepackage{amsmath,amssymb,amsfonts}
\usepackage{algorithmic}
\usepackage{graphicx}
\usepackage{multirow}
\usepackage{caption}
\usepackage{float}
\usepackage{hyperref}
\usepackage{tabularx}
\usepackage{makecell}
\usepackage{balance}
\usepackage{textcomp}
\usepackage{subcaption}
\usepackage{makecell}
\usepackage{balance}
\usepackage{dblfloatfix}

\usepackage[bottom]{footmisc}

\def\BibTeX{{\rm B\kern-.05em{\sc i\kern-.025em b}\kern-.08em
    T\kern-.1667em\lower.7ex\hbox{E}\kern-.125emX}}

\begin{document}
\history{Date of publication xxxx 00, 0000, date of current version xxxx 00, 0000.}
\doi{10.1109/ACCESS.2017.DOI}

\title{WWFedCBMIR: World-Wide Federated Content-Based Medical Image Retrieval }
\author{\uppercase{Zahra Tabatabaei}\authorrefmark{1,2},
\uppercase{Yuandou Wang}\authorrefmark{3}, \uppercase{Adrián Colomer}\authorrefmark{2,4}, \uppercase{Javier Oliver}\authorrefmark{1}, \uppercase{Zhiming Zhao}\authorrefmark{3}, \uppercase{Valery Naranjo}\authorrefmark{2}}

\address[1]{Dept. of Artificial Intelligence, Tyris Tech S.L.,Valencia, Spain.}
\address[2]{Instituto Universitario de Investigación en Tecnología Centrada en el Ser Humano, HUMAN-tech, Universitat Politècnica de València, Spain.}
\address[3]{Universiteit van Amsterdam, The Netherlands.}
\address[4]{ValgrAI – Valencian Graduate School and Research Network for Artificial Intelligence}
\tfootnote{This study is funded by European Union’s Horizon 2020 research and innovation program under the Marie Skłodowska-Curie grant agreement No. 860627 (CLARIFY Project).
The work of Adrián Colomer has been supported by the ValgrAI – Valencian Graduate School and Research Network for Artificial Intelligence \& Generalitat Valenciana and Universitat Politècnica de València (PAID-PD-22).}

\markboth
{Author \headeretal: Preparation of Papers for IEEE TRANSACTIONS and JOURNALS}
{Author \headeretal: Preparation of Papers for IEEE TRANSACTIONS and JOURNALS}

\corresp{Corresponding author: Zahra Tabatabaei (e-mail: elec.tabatabaei@gmail.com/ zahra.tabatabaei@tyris-software.com).}

\begin{abstract}

The paper proposes a Federated Content-Based Medical Image Retrieval (FedCBMIR) platform that utilizes Federated Learning (FL) to address the challenges of acquiring a diverse medical data set for training CBMIR models. CBMIR assists pathologists in diagnosing breast cancer more rapidly by identifying similar medical images and relevant patches in prior cases compared to traditional cancer detection methods. However, CBMIR in histopathology necessitates a pool of Whole Slide Images (WSIs) to train to extract an optimal embedding vector that leverages search engine performance, which may not be available in all centers. The strict regulations surrounding data sharing in medical data sets also hinder research and model development, making it difficult to collect a rich data set. The proposed FedCBMIR distributes the model to collaborative centers for training without sharing the data set, resulting in shorter training times than local training. FedCBMIR was evaluated in two experiments with three scenarios on BreaKHis and Camelyon17 (CAM17). The study shows that the FedCBMIR method increases the F1-Score (F1S) of each client to 98\%, 96\%, 94\%, and 97\% in the BreaKHis experiment with a generalized model of four magnifications and does so in 6.30 hours less time than total local training. FedCBMIR also achieves 98\% accuracy with CAM17 in 2.49 hours less training time than local training, demonstrating that our FedCBMIR is both fast and accurate for both pathologists and engineers. 
In addition, our FedCBMIR provides similar images with higher magnification for non-developed countries where participate in the worldwide FedCBMIR with developed countries to facilitate mitosis measuring in breast cancer diagnosis. We evaluate this scenario by scattering BreaKHis into four centers with different magnifications.

\end{abstract}


\titlepgskip=-15pt

\maketitle

\section{Introduction}
\label{sec:introduction}
\PARstart{B}{reast} cancer accounts for 25\% of all cancers in women worldwide. According to the American Cancer Society, a woman is diagnosed with breast cancer somewhere in the world every 14 seconds. In the year 2020, approximately 2.3 million women were diagnosed with breast cancer globally, and 685,000 lost their lives due to it. Histopathology is commonly used in the diagnosis and treatment of various diseases, including cancer. A biopsy, which is the removal of a small piece of tissue from the body, is usually required for histopathological examination. Typically, pathologists review the tissue to localize suspicious regions and, depending on the cancer type, they adjust their microscope to be able to analyze the tissue at various magnifications. Human error in histopathology refers to mistakes or inaccuracies made during the process of examining tissues or cells under a microscope. Some examples of human errors in histopathology include, Sampling errors, Processing errors, Technical errors, Interpretation errors, and Reporting errors. To minimize human errors in histopathology, it is essential to follow strict protocols and guidelines, perform regular quality control checks, and ensure that all personnel involved in the process are properly trained and competent. Quality assurance programs, such as peer review and proficiency testing, can also help to identify and correct errors.
Digital pathology can help pathologists to improve the accuracy and efficiency of cancer diagnosis, reduce the risk of errors, and enhance patient care. \cite{kim2012diagnostic} analyzed the accuracy of breast cancer diagnosis in 102 cases and found that there were diagnostic errors in 15.7\% of cases. The most common types of errors were misclassification of tumor type and misinterpretation of pathology slides.

Digital pathology is a technology that uses digital images of tissues and cells to aid in the diagnosis and management of diseases. This technology can help pathologists in several ways and reduce the risk of human error in cancer diagnosis. Deep Learning (DL) has revolutionized Computer-Aided Diagnosis (CAD) in digital pathology and opened the door to improve cancer diagnosis while decreasing the pathologist's workload. 

\begin{figure}[b]	
    \centering
    \includegraphics[width=0.45\textwidth]{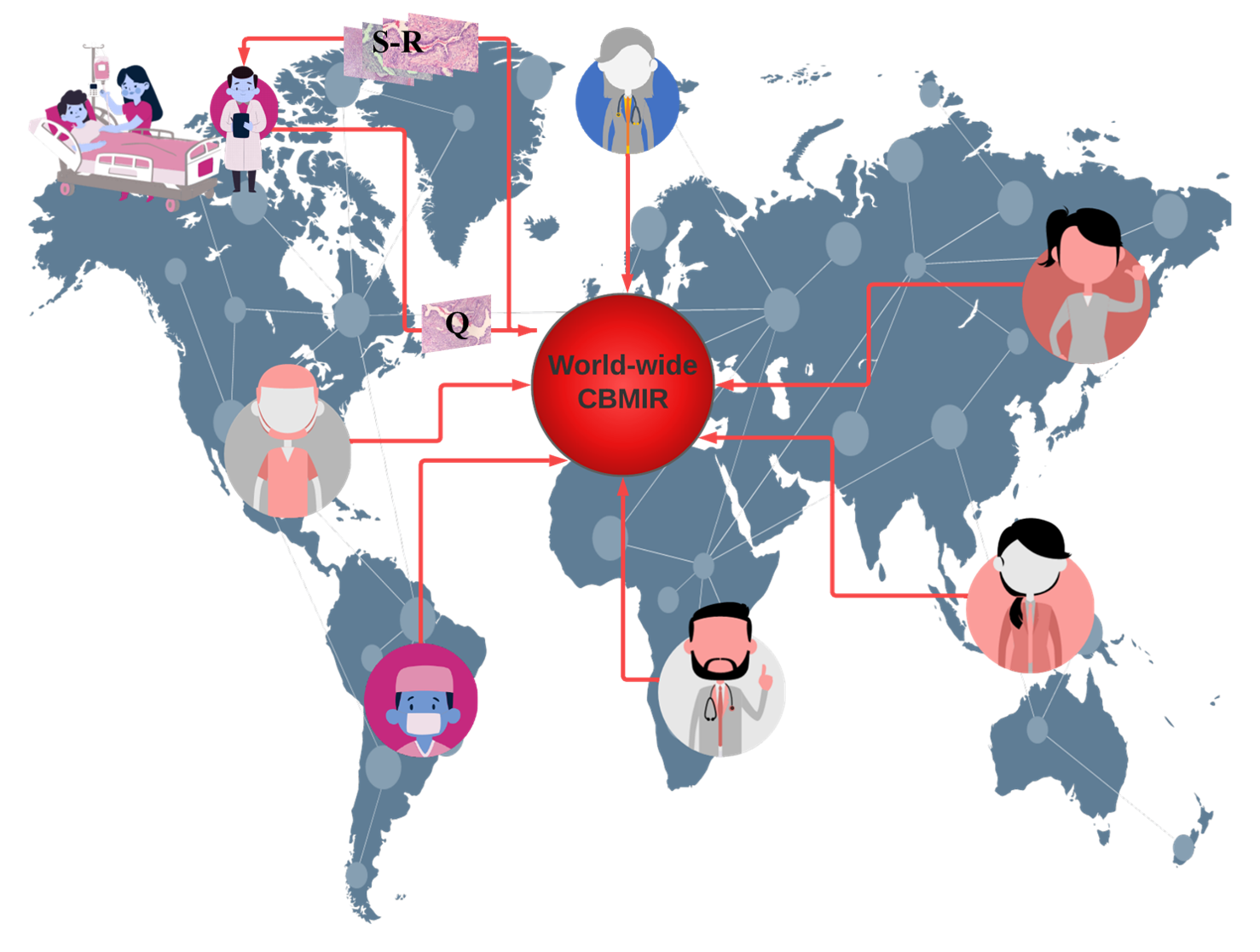}
    \caption{An overview of the use case of a worldwide CBMIR. 
    Pathologists send their Query (\textbf{Q}) to the worldwide CBMIR since they need a second opinion to make a more confident decision. Then, the model retrieved top \textit{K} similar images (\textbf{S-R}), and the pathologists can get a second opinion from whole over the world. }
    \label{fig:world-wide}
\end{figure}

Content-Based Medical Image Retrieval (CBMIR) is a recent DL-based methodology that allows for a quick and precise search in previously diagnosed and treated cases. In CBMIR, image features such as texture, shape, color, and intensity are extracted from the query image, and a similarity measure is applied to compare these features with those of the images in the database. The retrieved images are ranked according to their similarity to the query image, and the most relevant images are displayed to the user. Figure~\ref{fig:world-wide} shows how a worldwide CBMIR can provide unprecedented access to \textit{K} number of patches with the most similar patterns, allowing the pathologists to make a more confident diagnosis. By combining information from retrieved comparable instances with query, pathologists can enhance their initial diagnosis, obtain a better picture of the patient's prognosis, and ultimately come up with an appropriate final diagnostic interpretation. 

One of the advantages of CBMIR from the user's perspective is that it is not a completely black box for them. 

In other words, by receiving the output images, they can find similar patterns among them and with their query based on their knowledge. This provides more reliable information than a label for pathologists, which makes CBMIR more beneficial for pathologists than a classification.

A real-world scenario needs a global CBMIR, which demands a generalized data set with a variety of images of different quality, magnification, color, size, etc. The performance of CBMIR relies on a vast amount of data, which is difficult to collect in the medical field due to patient privacy and time costs. A promising CBMIR does have more chance to seek similar patterns in a pool of images in comparison to a limited data set. In order to create a vast centralized data set, DL experts need to transfer their Whole Slide Images (WSIs); however, these images are gigapixels, with high storage size for each. In addition to the challenges of transferring a heavy data set for DL experts, data protection and other regulatory obstacles on the medical side make it more challenging to create a sufficient data set. 

FL represents a possible solution to tackle this problem by collaboratively training DL models without transferring WSIs~\cite{mcmahan2017communication}. Multiple parties can safely co-train DL models in digital pathology using FL, achieving cutting-edge performance with privacy assurances~\cite{lu2022federated}. FL brings an opportunity to share the weights for multi-institutional training without sharing patient data and images. However, there are still some privacy risks since the training parameters and model weights are distributed among collaborators~\cite{sheller2020federated}. 
Our main contributions include as follows: 
\begin{itemize}
    \item Our research proposes a new approach called Federated Content-based Medical Image Retrieval (FedCBMIR) to address the challenges in traditional deep learning (DL) based image retrieval and federated learning (FL). 

    \item We conducted extensive experiments on FedCBMIR across different nations without sharing local data sets and compared the results with a locally trained CBMIR. Our results show that FedCBMIR outperforms traditional CBMIR while saving training time. 

    \item Our proposed approach allows pathologists to retrieve images at different magnifications from previous cases, which can benefit pathologists in non-developed countries with limited access to high-magnification scanning equipment.
    
    \item We decreased the time cost for model training and improved accuracy by using a joint data set. 
\end{itemize}

\section{Related work}{\label{Related work}}

In recent times, researchers have directed their attention toward both FL and CBMIR and have invested their efforts in exploring these domains. The subsequent subsections provide a succinct overview of some of these studies.

\subsection{Content-Based Medical Image Retrieval (CBMIR)}\label{Content Based Medical Image Retrieval (CBMIR)}
Content-Based Image Retrieval (CBMIR) has been a subject of extensive research since the advent of large-scale databases nearly two decades ago, as noted by Wang \cite{wang2022two}. Several studies have made significant contributions to this field. For example, Tabatabaei \cite{tabatabaei2022residual} achieved an accuracy rate of 84\% in CBMIR using the largest patch-annotated dataset in prostate cancer. Kalra \cite{kalra2020yottixel} proposed Yottixel, a method for representing The Cancer Genome Atlas Whole Slide Images (TCGA WSIs) compactly to facilitate millions of high-accuracy searches with low storage requirements in real-time. Conversely, Mehta \cite{5334811} proposed a CBMIR system for sub-images in high-resolution digital pathology images, utilizing scale-invariant feature extraction. Lowe \cite{lowe1999object} utilized Scale-Invariant Feature Transform (SIFT) to index sub-images and reported an 80\% accuracy rate for the top 5 retrieved images. Lowe's experiments were conducted on 50 ImmunohHistoChemistry (IHC) stained pathology images at eight different resolutions. Additionally, Hegde \cite{hegde2019similar} used a manually annotated dataset pre-trained on a Deep Neural Network (DNN) to achieve top 5 scores for patch-based CBMIR at different magnification levels.
The primary focus of recent studies has been on enhancing the performance of Content-Based Medical Image Retrieval (CBMIR) in different types of cancer; however, there are still several challenges that can impede its effectiveness. These challenges include data privacy, as medical data is confidential and subject to strict privacy regulations, making it arduous to share and access large data sets for model training. FL can alleviate this issue by facilitating distributed model training on local data without compromising privacy. Another challenge is data distribution, as medical data is frequently dispersed across numerous locations, making it difficult to train models on a centralized data set. FL can enable the training of models across multiple distributed data sets without aggregating the data in a central location. In addition, medical data sets can be heterogeneous, varying in terms of imaging modalities, quality, and annotation protocols, which can impede the development of robust and accurate models. FL can mitigate this challenge by allowing models to be trained on diverse data sets, improving their performance and generalization ability. Furthermore, medical data sets can be large and complex, necessitating significant computational resources to train models. FL can distribute the computational workload across multiple devices and locations, enhancing scalability and reducing training time.

\subsection{Federated Learning (FL)}

In recent years, FL has achieved impressive progress that enhances a wide adoption of DL from decentralized data~\cite{mcmahan2017communication,bonawitz2019towards, reddi2020adaptive}. FL is a distributed machine learning approach that can effectively handle decentralized data without raw data exchange to train a joint model by aggregating and distributing local training. Many existing algorithms can be adopted to aggregate updates from distributed clients. Typical examples include FederatedAveraging -\textit{viz} FedAvg~\cite{mcmahan2017communication}, and adaptive federated optimization methods~\cite{reddi2020adaptive}, e.g., FedAdagrad, FedYogi, and FedAdam. 
Some popular FL frameworks such as TensorFlow Federated (TFF)~\footnote{https://www.tensorflow.org/federated}, PySyft~\cite{ziller2021pysyft}, and Flower~\cite{beutel2020flower} have provided with a set of the robust set of tools for building privacy-preserving ML models. Besides, Jupyter Notebook-based tools such as~\cite{launet2023federating} also help simplify the FL setup and enable its deployment of a cross-country federated environment in only a few minutes. Daniel Truhn in~\cite{truhn2022encrypted} employed Homomorphic encryption to protect the model's performance while training by encrypting the weight updates before sharing them with the central server. Firas Khader in~\cite{khader2022medical} presented a technique of ``learnable synergy'', where the model only chooses pertinent interactions between data modalities and maintains an``internal memory'' of key information. Micah J. Sheller~\cite{sheller2020federated} investigated that FL among ten institutions is 99\% as efficient as those derived using centralized data. One recent work related to content-based image retrieval is introduced in~\cite{zhang2022flsir}, where FLSIR was proposed, and it enables secure image retrieval based on
FL and additive secret sharing. Nevertheless, it is not for clinical applications. The context of CBMIR and FL by literature study is still new, 
though the combination of these two techniques can significantly enhance healthcare outcomes by providing healthcare professionals with immediate access to precise and pertinent medical image data while ensuring patient privacy. 

The following sections will elaborate how the proposed FedCBMIR approach can revolutionize the way medical images are searched and utilized, leading to improved diagnoses and treatment plans.

\section{Federated Content-Based Medical Image Retrieval and its applications}
DL-model gives information that goes beyond the scope of human vision, and FL can expand it by removing national boundaries and connecting centers without regard for policy. This benefit can remedy the health care limitations due to the lack of facilities (staining materials, scanners, etc.) and experience (students, recently graduated pathologists, etc.). 

Hematoxylin and Eosin (H\&E), and IHC are two types of histopathological staining. H\&E has been popular for almost a century because it may indicate morphological changes~\cite{fischer2008hematoxylin}. In the common cancer diagnosis, IHC provides a high-resolution distribution and localization of specific cellular components within a tissue for pathologists who needs more details~\cite{ramos2005technical}. In some non-developed countries, the required IHC equipment is not available, which can increase the rate of uncertainty in the diagnosis report. 

CBMIR can be a valuable tool for new graduate pathologists in their training and professional practice, offering benefits such as improved education, decision-making, research, and time efficiency. This means CBMIR makes a platform for pathologists of all levels of expertise and experience to connect with one another and receive second opinions on their tissue. 


In order to address the challenges described above, we introduce a new approach called Federated Content-Based Medical Image Retrieval (FedCBMIR). This method allows for the effective management of decentralized medical images by utilizing local training for multiple tasks, while avoiding the need for raw data exchange. In this paper, we refer to this method as FedCBMIR, which is inspired by a great vision about a World-Wide Content-Based Medical Image Retrieval platform. A worldwide CBMIR provides a chance to have a more accurate diagnosis for less-developed countries without paying the cost of IHC. On this occasion, pathologists can rely on the labels provided by other pathologists in wealthy nations since they have higher confidence in their grading as a result of their facilities.

Image content is a key factor in the field of CBMIR; the more effective content is described, the more generalized the CBMIR framework will be. FL can give CBMIR a higher chance of generalizing its capabilities by accessing multi-central images from different hospitals.

CLoud ARtificial Intelligence For pathologY (CLARIFY) project \footnote{http://www.clarify-project.eu/} has a multi-institutional paradigm that includes nine institutions across five cities in three different countries and aims to provide an environment through the use of cloud computing and DL to facilitate knowledge sharing across institutions. In this work, according to the connections between different centers in CLARIFY, four centers (three universities and one company) in three cities in two countries gathered to mimic the real-world use case of FL in CBMIR. 

In this paper, we cope with the challenges of CBMIR with two different experiments and evaluate it in three scenarios. In our first experiment, we mimic a case of two centers that have different breast cancer WSIs in a completely different image preparation processes. This case occurs when two centers have a limited number of images, but they need a well-trained model to obtain a supportive idea on their query tissue. We assess this experiment on CAMELYON17 and BreaKHis at 40$\times$ magnification. Then, we extended our work with patches at different magnifications by feeding our FedCBMIR framework with BreakHis data set at 40$\times$, 100$\times$, 200$\times$, and 400$\times$ magnification. The magnification problem in WSI analysis is the subject of our second experiment.

The novelty of this work is providing well-trained models that can retrieve similar patches for each client in different countries. Regarding the use of FL in CBMIR, all clients with respect to data privacy can train the model with a limited number of patches and find similar patches to their queries more accurately than local training.

We validated the performance of CBMIR with FL on histopathological images using a custom-built Convolutional Auto Encoder (CAE) in a cross-institutional distributed environment. 
We used FL as a collaborative learning paradigm, in which the CAE can be trained across different centers and institutions without explicitly sharing data sets.

\section{Experiments}\label{sec: exp}
In this section, we introduce our FedCBMIR platform along with the training details and the two data sets used in our study.
\subsection{Materials}
\subsubsection{BreaKHis}
BreaKHis contains 7,909 histopathological images of breast tumor tissues that were provided by a collaboration with the P\&D Laboratory—Pathological Anatomy and Cytopathology, Parana, Brazil. This data set was collected from 82 patients at four magnifications (40$\times$, 100$\times$, 200$\times$, and 400$\times$) with 2,480 benign and 5,429 malignant cases. As can be understood in Table \ref{tab:break_info}, the number of images in benign and malignant cases is imbalanced. The most considerable portion of the data set belongs to the images at 100$\times$ magnification \footnote{https://www.kaggle.com/datasets/ambarish/breakhis}.
\begin{table}[htp!]
    \centering
    \caption{The distribution of BreakHis data set.}
    \begin{tabular}{|c|c|c|c|}
    \hline
     \textbf{Magnification} & \textbf{Benign} & \textbf{Malignant} & \textbf{Total}\\
    \hline\hline
          40$\times$ & 625 & 1370 & 1995 \\
         \hline
          100$\times$& 644 & 1437 &2081\\
         \hline
          200$\times$& 623 &1390 &2013 \\
         \hline
          400$\times$&588 & 1232 & 1820 \\
         \hline
         Total & 2480 & 5429 & 7909\\
         \hline
    \end{tabular}
  
    \label{tab:break_info}
\end{table}
\subsubsection{CAMELYON17 (CAM17)}

The database belonging to the CAMELYON17 (CAM17) challenge, as described by \cite{tellez2019quantifying}, is utilized for breast cancer metastasis detection in lymph node sections and will be used in classification experiments that involve the use of SCA in CN. CAM17 is composed of 1000 WSIs from 5 medical centers, but only the training set, which consists of 500 WSIs, was used due to the lack of annotations for the test WSIs. To do so, in this paper, images from four centers were used for training and validating the model, and the images of the 5th center were fed into the model as a test set. The data set includes 20 patients per center and five slides per patient, and pathologists annotated cancer regions only on 50 WSIs. Non-overlapping 224$\times$224 (at 40$\times$) pixel patches with at least 70\% tissue were used for experiments on this data set. In the experiments of this paper, the data set was considered as a binary data set, including Cancerous (annotated) and Non-Cancerous (not annotated). 

\subsection{Data distribution}
In the first experiment, Tyris Software (TY)\footnote{Spain, Valencia} and the Universiteit van Amsterdam (UvA) \footnote{Amsterdam, The Netherlands} have the CAM17 and BreaKHis 40$\times$, respectively (Table ~\ref{tab4:info_first_exp}).


In the second experiment, regarding mimicking the real-world data limitation, each institution (client) in this paper has BreakHis at only one magnification to train their model (Table~\ref{tab:info}). Universidad de Granada (UGR)\footnote{Spain, Granada}, TY, UvA, and Universidad Politécnica de Valencia (UPV)\footnote{Spain, Valencia} trained the custom-built CAE with BreakHis 40$\times$, 100$\times$, 200$\times$, and 400$\times$, respectively. 

\begin{table*}[htp!]
\setlength\tabcolsep{0pt}
\begin{center}
\centering
\caption{The initial experiment involved gathering data about each client, such as their location, the name of their center, the data associated with their data distribution, as well as the GPUs utilized for both training and searching tasks in each center.}
\label{tab4:info_first_exp}
\begin{tabularx}{0.98\textwidth}{|c|>{\centering\arraybackslash}X|>{\centering\arraybackslash}X|>{\centering\arraybackslash}X|>{\centering\arraybackslash}X|>{\centering\arraybackslash}X|}
\hline
\textbf{{Client}}&\textbf{{Region}}&\textbf{{Institution}}&\textbf{Data set} &\textbf{GPU type}\\
\cline{1-5} 
\hline\hline

                1 & Valencia, Spain             &   TY  & CAM17  & NVIDIA GeForce RTX 3090  \\ \hline 
                2 & Amsterdam, The Netherlands &   UvA & BreakHis 40$\times$  & NVIDIA Tesla T4  \\ \hline

\end{tabularx}
\end{center}
\end{table*}

\begin{figure}[b]
    \centering
    \includegraphics[width=0.45\textwidth]{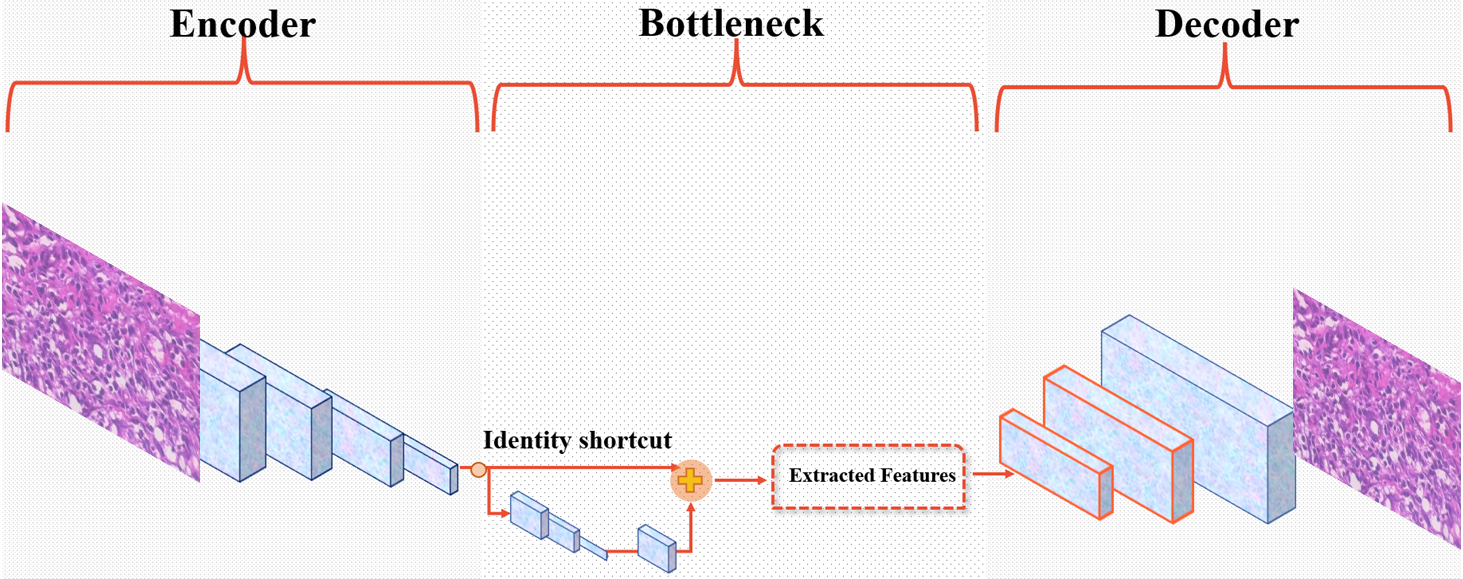}
    \caption{The structure of the custom-built CAE.}
    \label{fig:break_CAE}
\end{figure}

\begin{table*}[t]
\begin{center}
\caption{Collects information on each client in the second experiment, including their country and city, the name of the center, the related data due to the data distribution, and the GPUs used for training and search tasks in each center.}
\label{tab4:info}
\begin{tabular}{|c|c|c|c|c|c|}
\hline
\textbf{{Client}}&\textbf{{Region}}&\textbf{{Institution}}&\textbf{Magnification} &\textbf{GPU type}\\
\cline{1-5} 
\hline\hline

                1 & Granada, Spain              &   UGR & 40$\times $ & NVIDIA GeForce RTX 3090  \\ \hline 
                2 & Valencia, Spain             &   TY  & 100$\times $ & NVIDIA GeForce RTX 3090  \\ \hline 
                3 & Amsterdam, The Netherlands  &   UvA & 200$\times $ & NVIDIA Tesla T4  \\ \hline
                4 & Valencia, Spain             &   UPV & 400$\times $ & NVIDIA TITAN V \\ \hline
\end{tabular}\label{tab:info}
\end{center}
\end{table*}

\subsection{Distributed clients in training Convolutional Auto Encoder }
One of the most crucial elements of CBMIR that influences search engine results is the Feature Extractor (FE). The objective of content-based image search is to efficiently compare an extracted feature from a query image to every image in a database to identify the matches that are most similar.

Lack of annotated images and bias are two major challenges that need to be considered in integrating DL into cancer diagnosis. Three factors have the potential to identify bias in medical studies: data-driven, algorithmic, and human bias. To tackle these obstacles, a custom-built CAE is configured as the FE in this paper as a generative model where it is trained to reconstruct its input in an unsupervised way. Figure \ref{fig:break_CAE} shows its architecture with convolutional filters in the size of $[32, 64, 128, 256]$ in the encoder and, respectively, $[128, 64, 32, 3]$ in the decoder. In this custom-built CAE, a residual block with the filter size of $[64, 32, 1, 256]$ takes place between the encoder and decoder. This takes the originally extracted features from the backbone as its input and serves as a new feature map that contains the context relations between its feature input. It has a skip layer to jump over the layers to not only lead the model to converge faster and minimize the training errors but also boost the representation power and tackle the vanishing problem. Bottleneck delivers one feature vector with $200$ features ($F_i = \{f_1, f_2, f_3, ..., f_{200}\}$) per encoder input image \textit{i}. The model aims to achieve the lowest Mean Squared Error (MSE) by comparing Input (\textit{I}) and Output (\textit{O}) and is penalized if the reconstruction \textit{O} differs from \textit{I}. Once the unsupervised training is completed by removing the decoder part, a powerful automatic FE with a suitable output layer is available to extract the desired features.


\subsection{Local Training}
\begin{figure}[b!]
    \centering
    \includegraphics[width=0.45\textwidth]{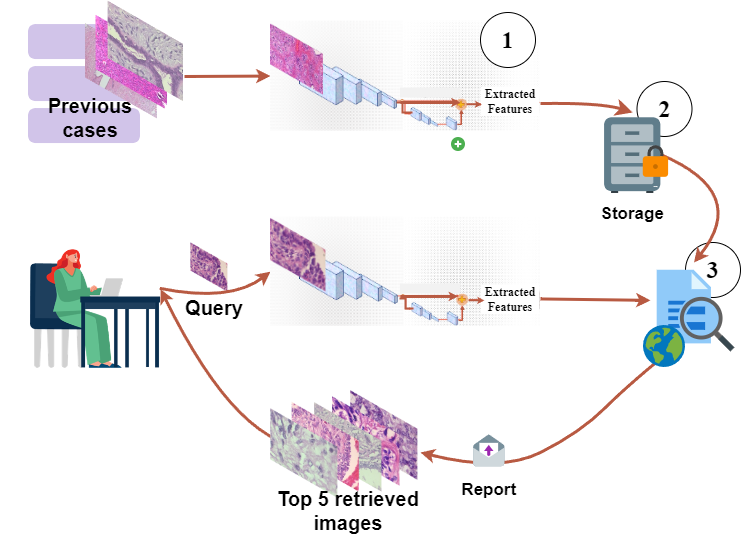}
    \caption{The pipeline of CBMIR. It contains three important sections such as 1) FE, 2) indexing and saving, and 3) similarity measure and search.}
    \label{fig:CBMIR}
\end{figure}

As it is mentioned above, by discarding the decoder, each institution has a well-trained FE to extract the desired discriminative features. Figure \ref{fig:CBMIR}.\footnote{BreaKHis images are used to plot the figure.}. explains the whole pipeline of the proposed CBMIR that each institution must follow to retrieve similar patches. Images in the training and validation set are fed into the FE to extract and save their features as the previous cases. All the $F_i$s are collected in a dictionary $D = [F_1, F_2, ... , F_n]$ in the middle of this figure.

From the interface perspective, pathologists upload their patch as a query image ($Q$) and expect to receive top \textit{K} similar patches. In practice, each $Q$ needs to represent as its feature vector $F_Q$ to the distance metrics in order to compare with the $F_i$s saved in $D$. To do so, as soon as the pathologists upload their $Q$, it is fed to the FE to extract $F_Q$ with $200$ features. Then, both $F_Q$ and the $F_i$s in $D$ are fed into the Euclidean distance to measure their similarity and deliver top \textit{K} similar patches.

\subsection{Federated learning configuration}
To achieve training DL models in a federated manner, 
in practice, we conducted experiments adopted in FedAvg and FedAdagrad strategies. In our cases, with a deep analysis, we found FedAvg performs better than FedAdagrad. Thus, this work adopts FedAvg to aggregate distributed updates from local clients: 
\begin{equation}
    \omega_{r+1} = \sum_{k=1}^K \frac{n_k}{n}\omega_{r+1}^k
\end{equation}
where $K$ indicates the number of clients, $r$ presents the communication round. For a client $k$ with $n_k$ samples, the local updates are arbitrary $\omega_{r+1}^k$. 
\begin{figure*}[b!]
\vspace{-5mm}
    \centering
    \begin{subfigure}[b]{\textwidth}
         \centering
        \includegraphics[scale=0.27]{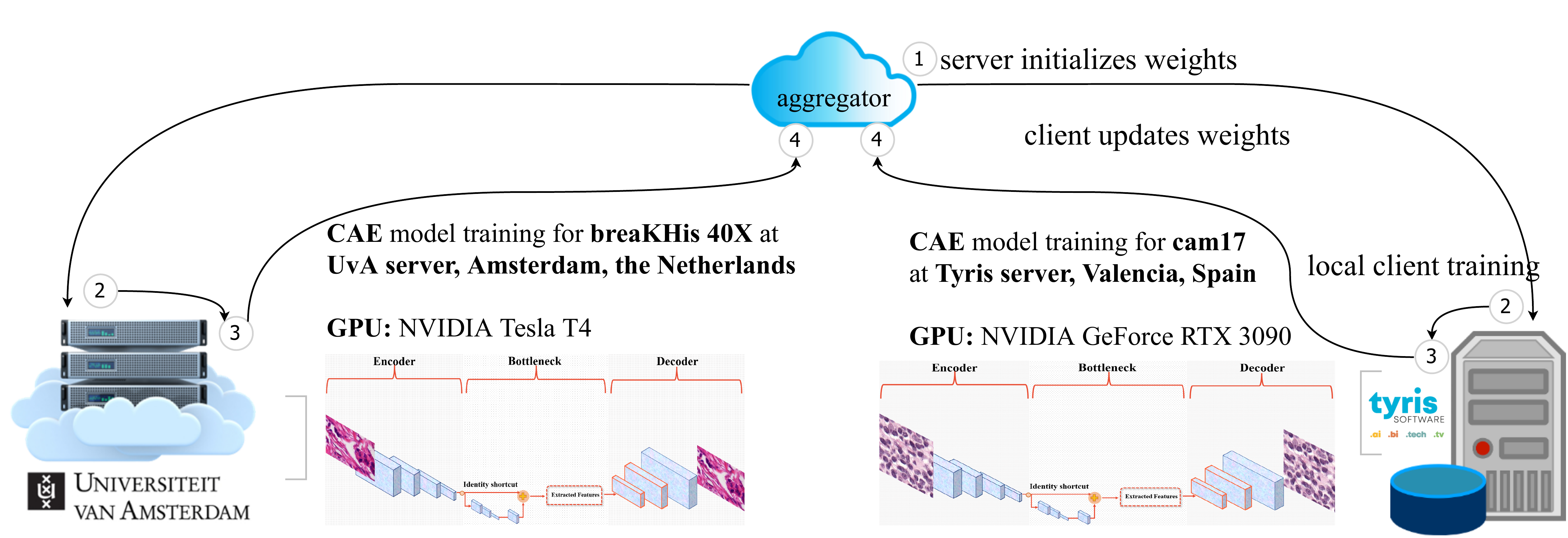}
        \caption{An overview of the FedCBMIR pipeline with two client training fed with breaKHis 40$\times$ and CAM17 data sets, respectively.}
        \label{fig: FedCBMIR2}
    \end{subfigure}
    \hfill
    \begin{subfigure}[b]{\textwidth}
         \centering
         \includegraphics[scale=0.27]{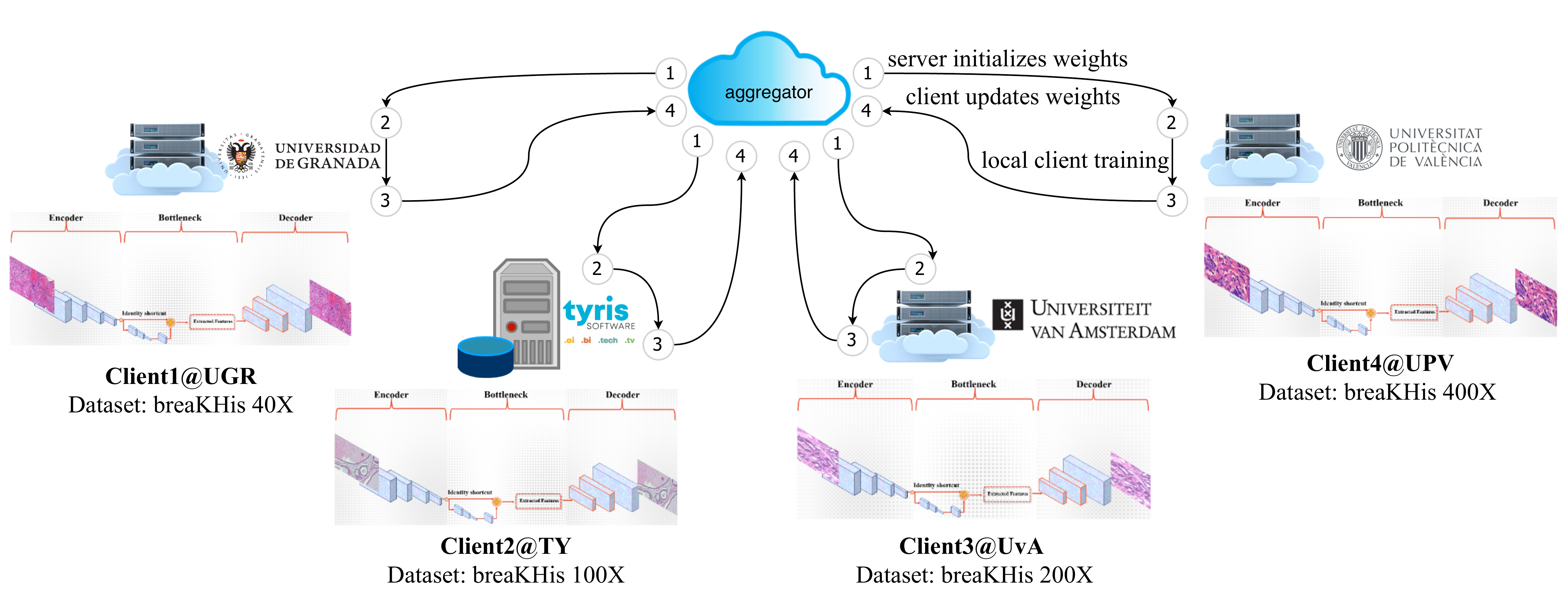}
         \caption{An overview of the FedCBMIR pipeline with four client training over clusters at universities and institutes in four different magnifications.}
         \label{fig: FedCBMIR4}
    \end{subfigure}
\caption{The FedCBMIR pipeline consists of four main steps. Step 1: the server initializes weights, and then sends to client for local training, step 2: client starts local training, step 3: client updates local weights to the server side, and step 4: the server side aggregates and updates the distributed weights.}
\end{figure*}

We apply FLOWER~\cite{beutel2020flower} as a primary framework to configure the FL experiments. We conduct two FL experiments as shown in Figure~\ref{fig: FedCBMIR2} and Figure~\ref{fig: FedCBMIR4}.
In the first experiment, it consists of two distributed training nodes located in TY and UvA, respectively (see in Figure~\ref{fig: FedCBMIR2}). We also extend FedCBMIR with more clients, as shown in Figure~\ref{fig: FedCBMIR4}. It is composed of four distributed nodes, each of them trains with BreaKHis data set in different magnifications. Table.~\ref{tab4:info} lists all four distributed processing nodes' information in the training phase.

\begin{table*}[htp!]
\centering
\caption{Provides a comparison in the test set between the performance of CBMIR and FedCBMIR in the first experiment as a result of aggregating CAM17 and BreaKHis 40$\times$. Hours and seconds, respectively, are used to measure the periods of training and searching. }
\begin{tabularx}{.98\textwidth}{|c|c|>{\centering\arraybackslash}X|c|c|c|c|c|}
\hline
                 \textbf{Center} &\textbf{Data} &\textbf{Model} &\textbf{Accuracy}  & \textbf{Precision} &\textbf{F1S} &\textbf{Training time}  &\textbf{Searching time}  \\ \hline
\multirow{3}{*}{\textbf{TY}} &\multirow{3}{*}{\textbf{CAM17}}   &  CBMIR & 0.96&0.96&0.96 &  8.7 h & 0.28 S \\ \cline{3-8} 
                  &  &FedCBMIR (Fedavg) &\textbf{0.98} & \textbf{0.97} & \textbf{0.98} &6.21 h &0.29 S \\ \cline{3-8}
                  & & FedCBMIR (FedAdagrad) & 0.98&0.97 &0.98 &7.92 h & 0.30 S\\
                  \hline\hline
\multirow{3}{*}{\textbf{UvA}} &\multirow{3}{*}{\textbf{BreaKHis40$\times$}}   & CBMIR &0.93&0.94&0.95 &9.33 h&0.018 S\\ \cline{3-8} 
                  &  &FedCBMIR (Fedavg)  & \textbf{0.98}  &\textbf{0.97}  &\textbf{0.98} & 6.59 h&0.024 S\\
                  \cline{3-8}
                  & &FedCBMIR (FedAdagrad) & 0.94&0.92 &0.96 &6.11 h & 0.04 S \\

                  \hline
                  
\end{tabularx}
\label{tab:CAM17_Break40X}
\end{table*}

\begin{figure*}[t!]
    \centering
    \includegraphics[width=1\textwidth]{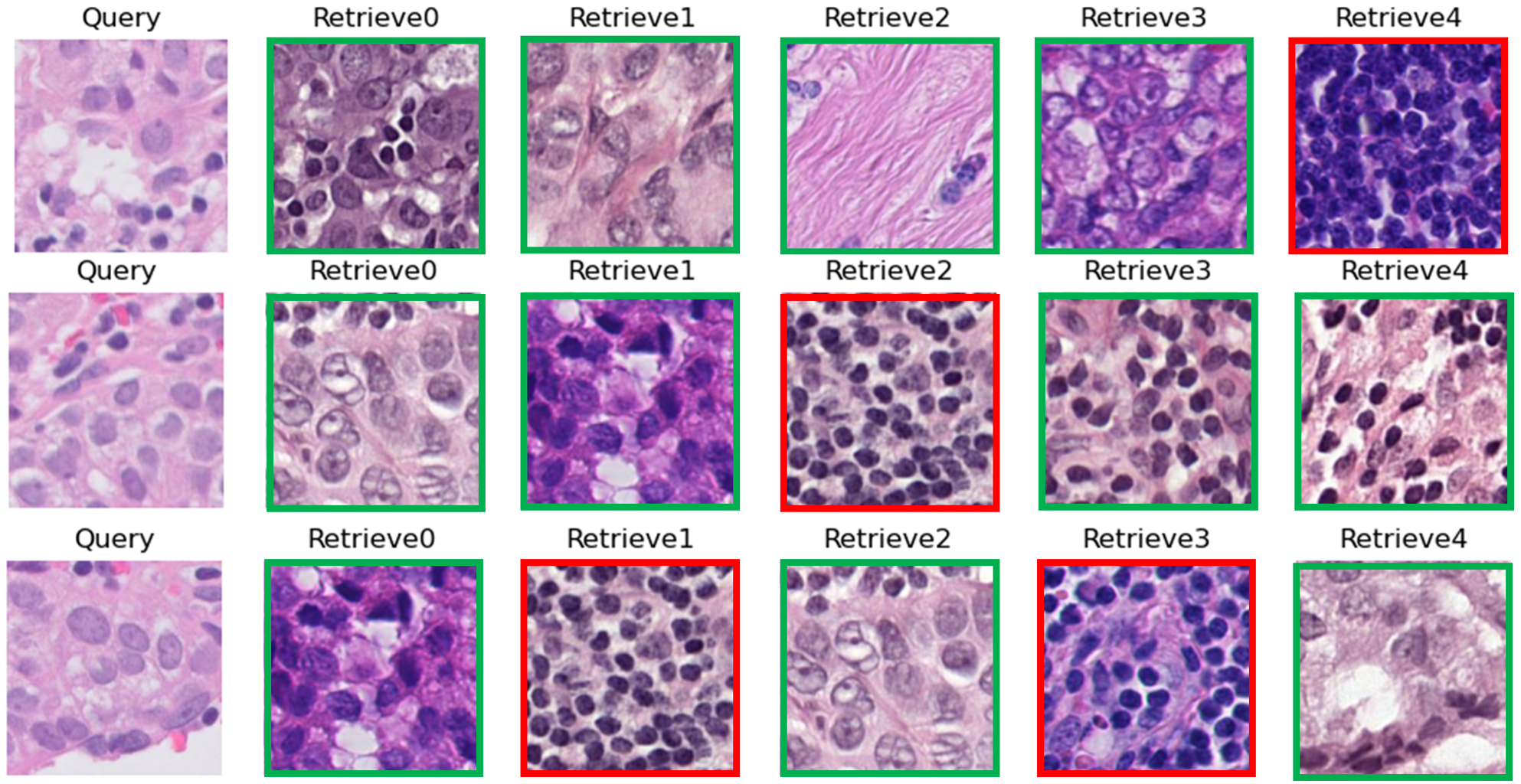}
    \caption{Shows three random queries from the fifth center of CAM17 (test set). Corresponding to each query top 5 images are shown from three other centers with the most similar patterns to the query. The green and red lines around the retrieved images explain the correct and wrong retrieved images.}
    \label{fig:cam17_top5}
\end{figure*}
\begin{figure}[b!]
\vspace{-5mm}
    \begin{minipage}{0.45\linewidth}
    \centering
    \begin{subfigure}[b]{\textwidth}
         \centering
        \includegraphics[scale=0.27]{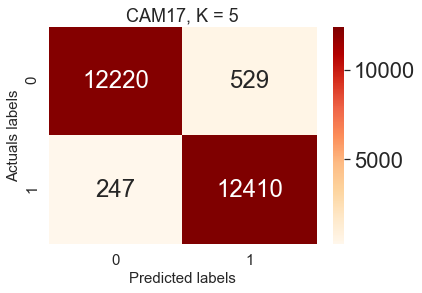}
        \caption{ }
        \label{fig: CAM17_5}
    \end{subfigure}
    \end{minipage}
    \hfill
    \begin{minipage}{0.45\linewidth}
    \begin{subfigure}[b]{\textwidth}
         \centering
         \includegraphics[scale=0.27]{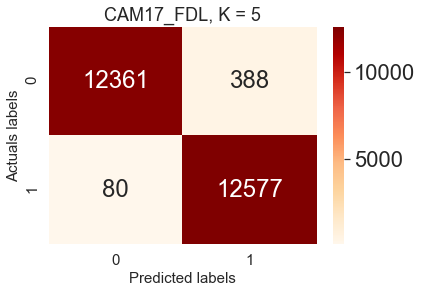}
         \caption{}
         \label{fig: CAM17_FDL_5}
    \end{subfigure}
    \end{minipage}
\caption{\ref{fig: CAM17_5}
shows the results of local training on CAM17 in the TY server. \ref{fig: CAM17_FDL_5} is the result of the searching task in CAM17 by applying the well-train FedCBMIR model from the first experiment.}\label{fig:CMs_CAM17}
\end{figure}




\section{Discussion and Results}\label{sec:res}
To allow for an adequate comparison of the model's performance, three metrics were selected: Accuracy (ACC), Precision, and F1Score (F1S), in addition to presenting the Confusion Matrix (CM). In this paper, regarding evaluating the proposed FedCBMIR, each of the images in the test set was considered a query. Across the entire training and validation set, the model is meant to detect similar patches.
\subsection{Results of the first experiments}
For this particular experiment, BreaKHis 40$\times$ and CAM17 data sets were aggregated to train the model. As a result, each center (UvA and TY) could develop a well-trained model to retrieve their respective images. The underlying assumption made in this experiment is that neither center had an agreement in place for sharing or accessing each other's images. Table~\ref{tab:CAM17_Break40X} provides a comprehensive view of the model and effectively demonstrates the credibility of the suggested FedCBMIR. To do this evaluation, the CAM images in center 5 were isolated from the images in the other four centers that were utilized for the training and validating task. Each image from center 5 serves as a query in the testing assignment, and the platform's function is to seek patches with a similar pattern from the other four centers. Table~\ref{tab:CAM17_Break40X} illustrates that the results of local training of CAM17 without aggregating with BreaKHis is less than the FedCBMIR as a result of well-training the model with data aggregation. The table indicates that FedCBMIR using the Fedavg approach achieved better results than CBMIR and FedAdagrad. As a result, Fedavg was selected as the aggregation technique for the subsequent experiments.

In terms of time and accuracy, local training of the CBMIR model on BreaKHis40$\times$ and CAM17 requires 9.33 and 8.7 hours, resulting in accuracies of 0.95\% and 96\%, respectively. However, FedCBMIR was trained more efficiently and achieved a higher accuracy level of 98\% in retrieving similar patches in CAM17, and 97\% accuracy for UvA, with a reduction of 2.49 and 2.74 hours in training time, respectively. Training time and accuracy are essential factors for DL scientists in building an optimal model, whereas accuracy and searching time are crucial for pathologists in retrieving similar patches. The results indicate that FL not only reduces training time, but also improves the accuracy and generalization ability of the model. The table shows that users in TY can obtain a second opinion with labels and similar patches in only 0.28 seconds per image, indicating the efficiency of FL in reducing searching time.

Figure~\ref{fig:cam17_top5} represents three random queries in the test set of CAM17 with their top 5 retrieved images among training and validation set. In order to have a more clear view of the obtained results, Figure~\ref{fig:CMs_CAM17} represents the comparison of image search results with two CMs in the test set of CAM17 as a result of local training (CBMIR) (~\ref{fig: CAM17_5}) and FedCBMIR (~\ref{fig: CAM17_FDL_5}).

\subsection{Results of the second experiments}

\begin{table}[h]
\centering
\caption{Obtained results of CBMIR on 40$\times$, 100$\times$, 200$\times$, and 400$\times$ at $\textit{K} = 5 $. We measure ACC, Precision, and F1S in the test set of each center at their corresponding magnification. Time is reported within an hour.}. 

\begin{tabularx}{.48\textwidth}{|c|X c c c c|}
\hline

\textbf{{Client}}&\textbf{ Model}&\textbf{Training time}&\textbf{Accuracy}&\textbf{Precision}&\textbf{F1S}\\ \hline
\multirow{2}{*}{1}  &CBMIR & 9.37 h& 0.95 & 0.93 &  0.96  \\ \cline{2-6} 
                  & FedCBMIR & 6.82 h & \textbf{0.97} & \textbf{0.96} &\textbf{0.98 }  \\ \hline\hline

\multirow{2}{*}{2} & CBMIR&  5.45 h &0.90 & 0.88 &0.94  \\ \cline{2-6} 
                  & FedCBMIR  & 5.78 h & \textbf{0.94}  & \textbf{0.92} & \textbf{0.96}   \\ \hline\hline

\multirow{2}{*}{3} & CBMIR & 8.59 h & 0.89 &0.87  &0.93  \\ \cline{2-6} 
                  & FedCBMIR & 6.65 h & \textbf{0.92} & \textbf{0.89} & \textbf{0.94} \\ \hline\hline

\multirow{2}{*}{4} & CBMIR & 8.95 h &0.92  &0.89  &0.94  \\ \cline{2-6} 
                  & FedCBMIR &6.83 h  & \textbf{0.96} & \textbf{0.94} &\textbf{0.97}  \\ \hline
\end{tabularx}\label{tab:break4magnifications}
\end{table}
\begin{figure*}[b!]
\centering
\begin{tabular}{cccc}
\centering
\includegraphics[width=0.23\textwidth]{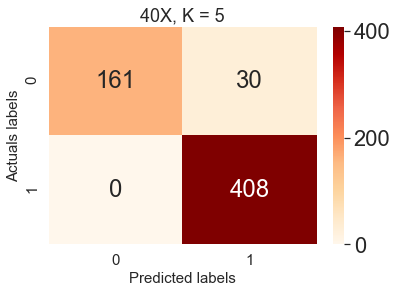} &
\includegraphics[width=0.23\textwidth]{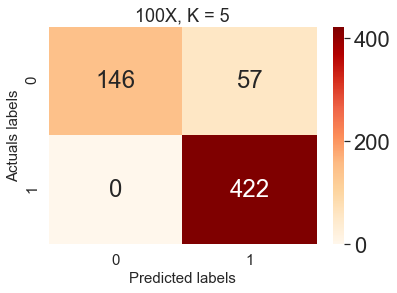} &
\includegraphics[width=0.23\textwidth]{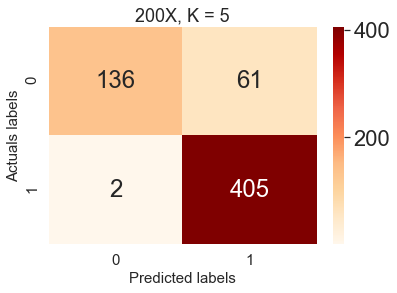} &
\includegraphics[width=0.23\textwidth]{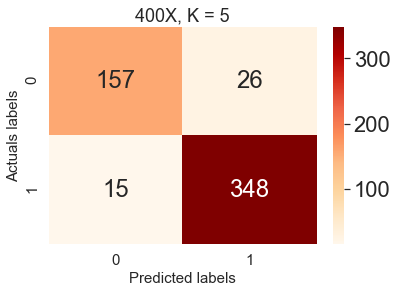} \\
\textbf{(a). 40$\times$}  & \textbf{(b). 100$\times$} & \textbf{(c). 200$\times$}  & \textbf{(d). 400$\times$}\\[6pt]
\end{tabular}
\begin{tabular}{cccc}
\centering
\includegraphics[width=0.23\textwidth]{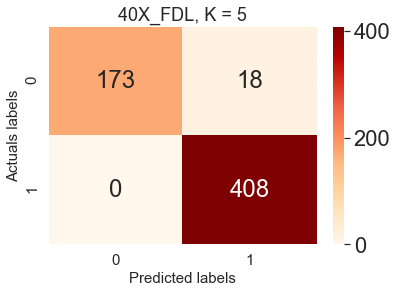} &
\includegraphics[width=0.23\textwidth]{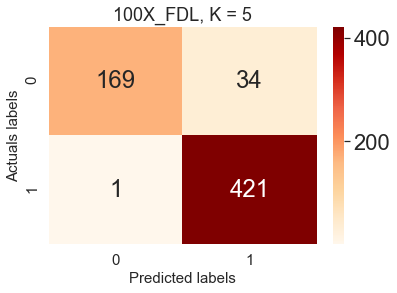} &
\includegraphics[width=0.23\textwidth]{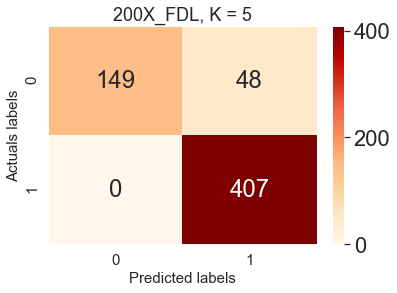} &
\includegraphics[width=0.23\textwidth]{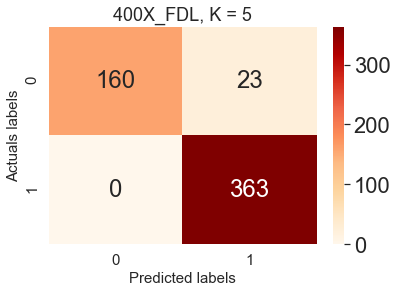} \\
\textbf{(e). 40$\times$}  & \textbf{(f). 100$\times$} & \textbf{(g). 200$\times$}& \textbf{(h). 400$\times$} \\[6pt]
\end{tabular}
\caption{ \textbf{(a)}-\textbf{(d)} show the CMs as a result of local training and searching at the same magnification. \textbf{e}-\textbf{h} are the CMs of FL models. The reported results are with top \textit{K} retrieved images. In all CMs, "0" and "1" indicate "\textbf{Benign} and "\textbf{Malignant}", respectively.  "\textbf{Actual labels}" and "\textbf{Predicted labels}" corresponds to the query and the retrieved labels, accordingly. }
\label{fig:CMs}
\end{figure*}

In the second experiment, the performance evaluation of the proposed framework was conducted using two distinct scenarios. The first scenario, named \textbf{\textit{Sen1}}, assumed that the centers did not have access to images from other centers, and were only allowed to share the model weights during the training phase. This scenario was designed to test the performance of the framework when the participating centers faced technical limitations in sharing large amounts of medical imaging data. In this scenario, each center had to train their model on their local data, and the models' weights were shared with other centers. Then, the weights were combined and trained using the entire data set from all participating centers. Finally, the model was evaluated on each center's local test set to assess its performance.

The second scenario, named \textbf{\textit{Sen2}}, assumed that the participating centers were only authorized to access comparable cases from the images located in other centers. This scenario was designed to evaluate the framework's performance when the participating centers could not share large files due to the difficulty of transferring them. In this scenario, each center trained their model on their local data, and a subset of the data from other centers was used to fine-tune the model. The fine-tuned models were evaluated on each center's local test set to assess their performance.

\textbf{\textit{Sen1}} mirrors the situation where centers can only obtain patches that are similar to their $Q$ at the same magnification. Because there is no explicit agreement among the centers, the model is obliged to search for similar cases in a few cases at that particular magnification. 


\begin{figure*}[b!]
    \centering
    \includegraphics[width=1\textwidth]{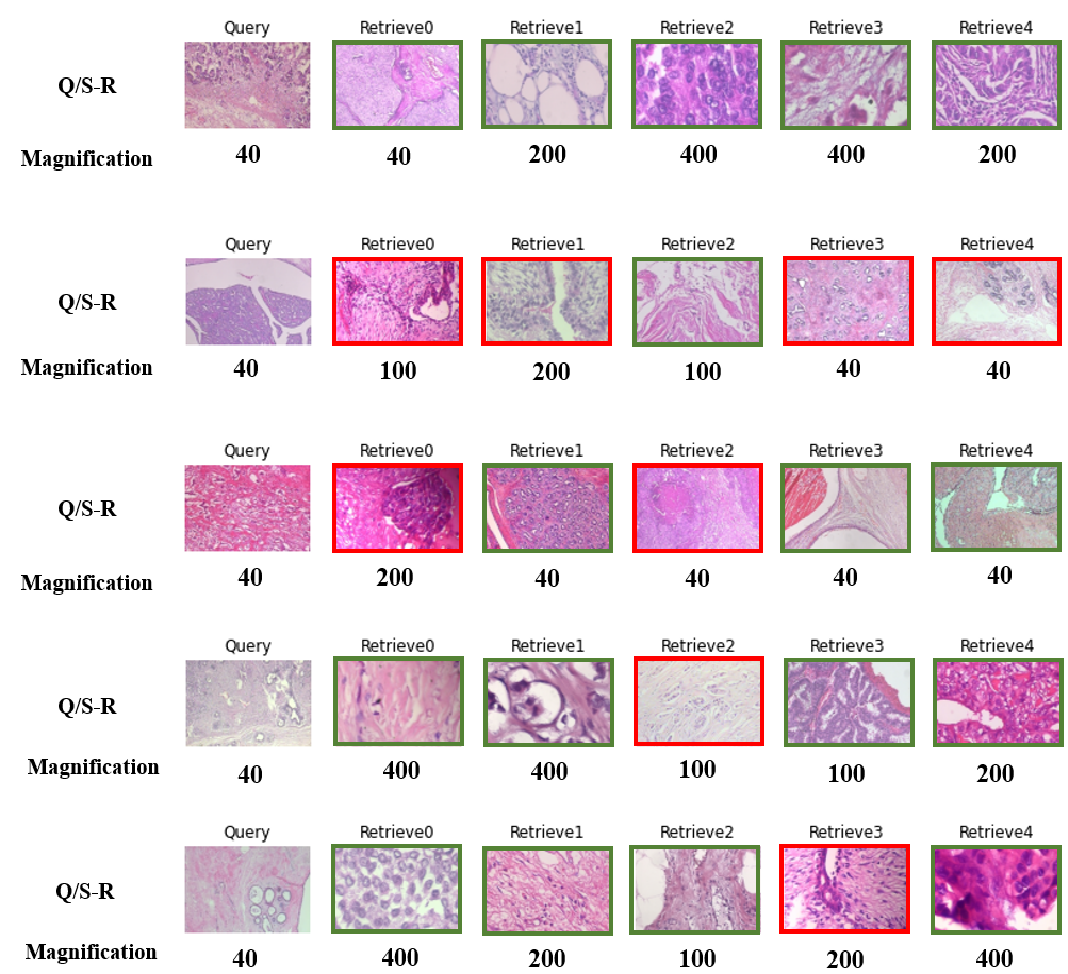}
    \caption{Shows five lines of random histopathological WSIs with their magnifications. The first column is the query, and the following five columns show the retrieved images. This figure brings a proper overview on \textbf{\textit{Sen2}}. For each query, the top 5 retrieved images are shown in green and red borders. The retrieved image with the same and different labels as the query is indicated by the green and red borders, accordingly. }
    \label{fig:search_Break_all_magnifications_4ex}
\end{figure*}

Table \ref{tab:break4magnifications} summarizes the results of the proposed FedCBMIR on the BreaKHis data set at all four magnifications. The table shows the accuracy and precision of the retrieved images at each magnification, achieved by each center after training their models for 300 epochs within their server and without using FL. The highest accuracy of 95\% for the retrieved images at 40$\times$ magnification was achieved by client one at UGR in 9.37 hours, while client 3 spent 8.59 hours to achieve a minimum accuracy of 89\% and precision of 87\%, which is the lowest among all the centers.

It is noteworthy that each center spent a total of 32.36 hours training four separate models with their limited data set. However, using FL, these four models were trained 6.28 hours faster (26.08 hours) with the entire training set of BreaKHis at its four magnifications. These results show that the proposed FedCBMIR method effectively improves the accuracy and efficiency of the CBMIR system in the medical field. It allows multiple centers with limited data sets to collaborate and train a single model on a larger data set, thereby achieving higher accuracy and precision in a shorter time. Moreover, the proposed method enables pathologists to access a user-friendly platform that can provide faster and more accurate solutions to their problems.

The performance evaluation of the proposed framework in the test set was compared with local training CBMIR and FedCBMIR, as shown in Figure~\ref{fig:CMs}. Each CM is associated with a specific magnification and reports the top 5 accuracy using \textbf{\textit{Sen1}} in its search stage. The results of the \textbf{\textit{Sen1}} scenario in the test set are presented in Figure~\ref{fig:CMs}, where each center receives the top 5 images on average in 13.84 seconds. The proposed approach provides an efficient and accurate search engine, which provides a user-friendly platform for pathologists to address their issues faster and easier.

It is important to highlight that the proposed approach surpasses both local training CBMIR and FedCBMIR, offering higher accuracy and efficiency. Furthermore, the use of FedCBMIR allowed the centers to train their models with their limited datasets while achieving an overall better performance by utilizing the entire training set of BreaKHis at all four magnifications. As demonstrated in Table~\ref{tab:break4magnifications}, using the proposed approach, the four models were trained 6.28 hours faster, thereby reducing the total training time from 32.36 hours to 26.08 hours. This reduction in training time is particularly significant for large datasets and can facilitate more rapid and accurate diagnoses and treatments of cancers.

The proposed approach, \textbf{\textit{Sen2}}, provides an efficient solution for pathologists to consult with other pathologists by accessing images at different magnifications, which can reduce inter-observer variability. In contrast to traditional CBMIR, \textbf{\textit{Sen2}} allows pathologists to retrieve similar cases at all four magnifications, not just from the same magnification as their query ($Q$). However, sharing images with a single server is not feasible due to storage and privacy concerns. To address this issue, the proposed FedCBMIR can retrieve similar patches at the same and higher magnifications. 
\begin{table}[h!]
\begin{center}
\caption{Illustrates the ACC, Precision, and F1S for the second scenario of the second experiment with \textit{K} = 5.}
\label{tab3:sen2}
\begin{tabular}{|c|c|c|c|c|c|}
\hline
\textbf{{Client}}&\textbf{Accuracy}&\textbf{Precision}&\textbf{F1S}\\
\cline{1-4} 
\hline\hline

\hline
                1  & 0.94 & 0.92 &   0.95 \\
\hline 
                2  & 0.95 & 0.93 &0.96\\
\hline 
                3  & 0.95 & 0.93 &  0.96  \\
\hline
                4 & 0.95 &  0.92&  0.96 \\
\hline
\end{tabular}
\end{center}
\end{table}
Table \ref{tab3:sen2} proves that the proposed FedCBMIR is highly robust to receive a query at a specific magnification and retrieve top 5 similar patches at all four magnifications. Each client fed the test set at the corresponding magnification and received top 5 retrieved patches at all four magnifications.
The results of feeding the model with five random queries at 40$\times$ magnification by following (\textbf{\textit{Sen2}}) are presented in Figure~\ref{fig:search_Break_all_magnifications_4ex}. The selected magnification of 40$\times$ is optimal because it is the lowest magnification in the dataset, and it is easier to measure the number of mitoses in images with higher magnifications. By feeding the model with images at 40$\times$, pathologists can receive top 5 similar images at 40$\times$, 100$\times$, 200$\times$, and 400$\times$, which can significantly reduce the time and effort required to obtain a second opinion. The proposed approach has the potential to improve the speed and accuracy of cancer diagnosis and treatment. As such, it can serve as a user-friendly platform for pathologists to more efficiently address their concerns. Furthermore, it has the potential to be a valuable tool for telepathology in the future.

One of the challenges in collecting WSIs for use in DLg models is the variability in color distribution due to differences in the staining material used across different centers and over time. This variability can have a significant impact on the accuracy and reliability of DL models. However, an important finding from the results shown in Figure~\ref{fig:search_Break_all_magnifications_4ex} is that the proposed approach, \textbf{\textit{Sen2}}, is not affected by differences in color distribution resulting from the staining process at different hospitals. This is a noteworthy result, as it indicates that the proposed approach can effectively overcome one of the major challenges associated with collecting and utilizing WSIs in telepathology. By eliminating the impact of color distribution variability, \textbf{\textit{Sen2}} provides a more robust and reliable platform for pathologists to obtain accurate and consistent diagnoses, regardless of the specific staining materials used at different centers.

 The availability of advanced technology, such as high-resolution scanners, is not always guaranteed in every part of the world. In this regard, the proposed \textbf{\textit{Sen2}} approach can serve as an important tool for pathologists in developing nations to overcome the limitations of their scanners by enabling them to access tissue images at higher magnifications. FedCBMIR can facilitate cross-border collaborations, where pathologists from different regions can share their knowledge and expertise by analyzing similar patches at higher magnifications. Therefore, the proposed approach can contribute significantly to improving the accuracy and speed of disease diagnosis, particularly in regions where access to advanced technology is limited. In this way, \textbf{\textit{Sen2}} has the potential to bridge the gap in healthcare and provide a more equitable and accessible healthcare system for all.

All the experimental results in both experiments and scenarios have verified that the proposed FedCBMIR has covered both concerns of DL scientists and pathologists with a fast-trained and accurate CBMIR, which is more generalized.

\section{Future work}\label{sec:future}

To further enhance the performance of Federated CBMIR for breast cancer diagnosis, it may be worthwhile to explore the use of additional data sets. This could include larger data sets with a greater number of labeled images, as well as data sets that encompass a wider range of malignancy levels and tumor subtypes. By incorporating these data sets into the FL process, it may be possible to improve the accuracy and robustness of the model.

In addition to expanding the data sets used in Federated CBMIR, it may also be valuable to incorporate other types of clinical data into the system. Patient demographic information and clinical history could provide additional context and help to further refine the diagnostic process. Exploring the integration of these types of data could be a promising avenue for future research.
\section{Conclusion}\label{sec: conclusion}

The present study proposes a FedCBMIR approach that addresses two significant challenges in digital pathology faced by pathologists and engineers. By retrieving the top five similar images in a short amount of time, the proposed method reduces the workload of pathologists and decreases the time and cost associated with developing a high-performing DL-based method.

To evaluate the proposed approach, two experiments were conducted with three different scenarios. Experiment 1 aimed to provide a generalized model with CAM17 and BreaKHis 40$\times$ for centers that do not have enough images to train a model effectively. The results for the CAM17 dataset showed a 97\% precision rate with a 29-second retrieval time for each query and a training time of 6.21 hours.

Experiment 2 comprised two scenarios: Sen1, where image centers are not in agreement for sharing images, and Sen2, where images are delivered in different magnifications for centers that lack the equipment to scan tissues at higher magnifications. The proposed method reached 98\%, 96\%, 94\%, and 97\% F1S for each center in Sen1. In Sen2, the BreaKHis dataset was distributed across four centers, resulting in accuracy rates of 97\%, 94\%, 92\%, and 96\% for pathologists at magnifications of 40$\times$, 100$\times$, 200$\times$, and 400$\times$, respectively. The average retrieval time was 13.84 seconds, and the well-trained model required 6.30 fewer hours to train.

Overall, this work offers a promising platform for medical centers to reduce the workload of pathologists by decreasing training time and increasing accuracy when compared to traditional CBMIR methods.

\section*{CRediT a
uthorship contribution statement}
\textbf{Zahra Tabatabaei:} Writing, Methodology, Analyzing, Reviewing \& editing, Formal analysis,  visualization.

\noindent \textbf{Yuandou Wang:}  Conceptualization, Methodology, Investigation, Formal analysis, Writing - Original Draft in sections related to federated learning, Writing - Review \& Editing. 

\noindent \textbf{Adrián Colomer:} Review

\noindent \textbf{Javier Oliver:} Review

\noindent \textbf{Zhiming Zhao:} Review

\noindent \textbf{Valery Naranjo:} Review

\section*{Acknowledgment }
We would like to express our gratitude to all who have contributed to the completion of this research paper. We thank Laëtitia Launet for her valuable assistance. We also acknowledge the University of Granada (UGR), especially Prof. Rafael Molina and Prof. Javier Mateos, for granting us access to their high-performance computing resources, including the NVIDIA GeForce RTX 3090 GPU, which greatly accelerated our deep learning computations and enabled us to achieve state-of-the-art results in our research. We are thankful for their support.

\section*{Abbreviations}
\begin{itemize}
\item Accuracy: ACC
\item CLoud ARtificial Intelligence For pathologY: CLARIFY
\item Computer-Aided Diagnosis: CAD
\item Confusion Matrix: CM
\item Content-Based Medical Image Retrieval: CBMIR
\item Convolutional Auto Encoder: CAE
\item Deep Learning: DL
\item Federated Content-Based Medical Image Retrieval: FedCBMIR
\item Federated Learning: FL
\item Feature Extractor: FE
\item F1Score: F1S
\item Hematoxylin and Eosin: H\&E
\item Immunohistochemistry: IHC
\item Tensorflow Federated: TFF
\item Whole Slide Images: WSIs
\end{itemize}

\bibliography{refs} \bibliographystyle{unsrt}

\EOD

\end{document}